\documentclass[sigconf]{aamas} 

\usepackage{balance} 

\setcopyright{acmcopyright}
\acmConference[AAMAS '24]{Working Paper 008}
{UK}{Simudyne}
\acmDOI{}
\acmPrice{}
\acmISBN{}

\acmSubmissionID{206}

\title[]{Agent-based Liquidity Risk Modelling for Financial Markets}

\author{Perukrishnen~Vytelingum}
\affiliation{
  \institution{Simudyne}
  \country{United~Kingdom}
}

\author{Rory~Baggott}
\affiliation{
  \institution{Simudyne}
  \country{United~Kingdom}
}

\author{Namid~Stillman}
\affiliation{
  \institution{Simudyne}
  \city{}
  \country{United~Kingdom}
}

\author{Jianfei~Zhang}
\affiliation{
  \institution{Hong Kong Exchanges and Clearing Limited}
  \country{Hong~Kong}
}

\author{Dingqiu~Zhu}
\affiliation{
  \institution{Hong Kong Exchanges and Clearing Limited}
  \country{Hong~Kong}
}

\author{Tao Chen}
\affiliation{
  \institution{Hong Kong Exchanges and Clearing Limited}
 \country{Hong~Kong}
}

\author{Justin~Lyon}
\affiliation{
  \institution{Simudyne}
  \country{United~Kingdom}
}

\begin{abstract}
In this paper, we describe a novel agent-based approach for modelling the transaction cost of buying or selling an asset in financial markets, e.g., to liquidate a large position as a result of a margin call to meet financial obligations. The simple act of buying or selling in the market causes a price impact and there is a cost described as liquidity risk. For example, when selling a large order, there is market slippage -- each successive trade will execute at the same or worse price. When the market adjusts to the new information revealed by the execution of such a large order, we observe in the data a permanent price impact that can be attributed to the change in the fundamental value as market participants reassess the value of the asset. In our ABM model, we introduce a novel mechanism where traders assume orderflow is informed and each trade reveals some information about the value of the asset, and traders update their belief of the fundamental value for every trade. The result is emergent, realistic price impact without oversimplifying the problem as most stylised models do, but within a realistic framework that models the exchange with its protocols, its limit orderbook and its auction mechanism and that can calculate the transaction cost of any execution strategy without limitation. Our stochastic ABM model calculates the costs and uncertainties of buying and selling in a market by running Monte-Carlo simulations, for a better understanding of liquidity risk and can be used to optimise for optimal execution under liquidity risk. We demonstrate its practical application in the real world by calculating the liquidity risk for the Hang-Seng Futures Index.
\end{abstract}

\keywords{Continuous Double Auction, Zero-Intelligence Model, Market Impact Model, Trading Strategy, Liquidity Risk, Optimal Execution}

\newcommand{\BibTeX}{\rm B\kern-.05em{\sc i\kern-.025em b}\kern-.08em\TeX}

\begin{document}


\pagestyle{fancy}
\fancyhead{}


\maketitle 


\section{Introduction}
Liquidity risk management assesses the risk that a financial institution will not be able to meet its financial obligations, e.g., if a trader does not have enough cash, collateral or funding. It is an important part of risk management for all financial institutions. For example, banks are required to hold cash or collateral to meet their daily cash flow needs and to understand the liability of large positions in the market. Exchanges need to calculate the margin traders are required to post when trading derivatives and to manage close-out activities in the event of a default. If a position needs to be unwound and liquidated in the market, the cost of execution can be significant, particularly when trying to liquidate large positions within a short horizon. This work takes a novel approach to the difficult problem of evaluating how size and execution horizon affect price impact using an agent-based modelling framework.

\begin{figure}[h]
    \centering
    \includegraphics[width=1\linewidth]{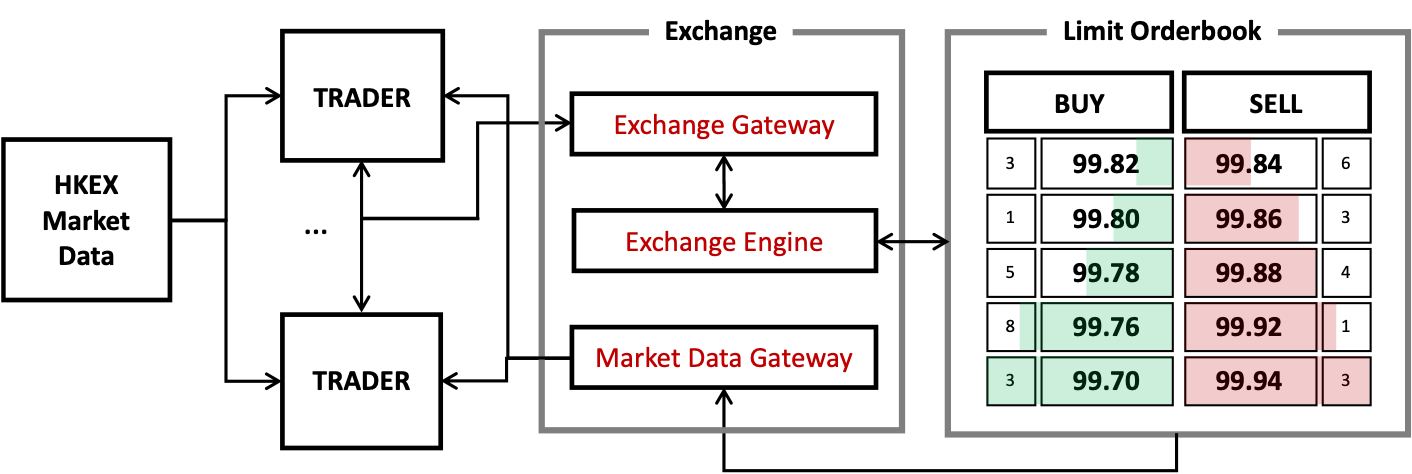}
    \caption{The market simulation framework with market data input, virtual trading agents, the exchange with its gateways, engine and protocols and the buy and sell limit orderbooks which are managed by an auction mechanism.}
    \label{fig:market_sim}
\end{figure}

Market behaviour has been studied from many perspectives, including stochastic theories \cite{Cont2019} and statistical models of limit orderbooks (LOBs), theoretical equilibrium models and so on; however, most of them have limitations and simplifications. More recently, agent-based models have also shown promise at reproducing market behaviour \cite{vyetrenko2020get,wang2017spoofing, cliff1997zero, gode1993allocative, bookstaber2018agent, mcgroarty2019high}. In an agent-based modelling (ABM) approach, we can model the problem without oversimplifying as a stylised model would. An ABM explicitly models the market as a discrete-step simulation with the mechanics of a Continuous Double Auction (CDA) of an exchange and trading agents with different trading strategies as we would find in a real market as shown in Figure \ref{fig:market_sim}. The ability of ABMs to capture the heterogeneity of trading behaviours and the structural complexity of an exchange has been shown to be effective in recreating the well-known stylized facts in financial markets, such as fat-tailed distribution of returns, volatility clustering or auto-correlation in returns \cite{PottersBouchaud_statprop}. Yet, a drawback of ABMs is in their modelling complexity which often allows only for a qualitative treatment of the results. 

This leads to the idea of a minimally complex ABM for explanability while reproducing the liquidity and volatility patterns of real markets. We begin with the so called Zero-Intelligence (ZI) model, which despite its simple and random trader behaviours, is able to produce the certain statistical properties of the market, e.g., fat tailed distributions of returns \cite{bookstaber2018agent}. Here, it is not only the trader behaviours that define market dynamics but the rules of the auction mechanism itself. A basic ZI trader submits limit and market orders at constant rates and orders on the LOB are cancelled at a constant rate. The price of a limit order is drawn from an exponential distribution. Importantly this ZI model can be calibrated to the historic data. While it has a well-understood behaviour and is able to capture some of the market features, the ZI model does not meet the requirement for a high-fidelity market simulator, but our ABM builds on similar principles.

We can breakdown our ABM into two parts. First, we have the structure, i.e., the rules of the model, here, a set of known mathematical equations and algorithms that define the mechanics of the exchange matching engine and its protocols. Against these rules, we have the behaviour of the agents with trading and execution strategies. The financial market is a complex system, and the challenge is to build a model with minimally complex behaviours that can be calibrated to market data. Across the different models described in the literature, we observe that value, trend and noise trading strategies are sufficient to capture the stylised behaviours of the market \cite{mcgroarty2019high,farmer2005predictive,wang2017spoofing} and we model these behaviours in this paper. Furthermore, we propose a mechanism where traders assume orderflow is informed and update their belief of the fundamental value as new information is revealed whenever a trade occurs. 

Agent-based modelling allows us to model behaviours of individual agents, including an agent that executes an order given an execution strategy, size and horizon. In the real world, a trader does not execute a large order (also referred to as a meta-order) immediately, but instead slices the order into many smaller ones. Intuitively, we might expect the price impact of the smaller orders to scale linearly, but the data describes a square-root law. In our ABM, as the meta-order is executed, we will empirically demonstrate in Section~\ref{sec:res} how we reproduce the square-root law and how transient and permanent market impact emerge rather than be explicitly coded into our model. The outcome is a realistic transaction cost model that can calculate liquidity risk in a novel way by breaking down costs to market risk and market impact. The cost is calculated as an implementation shortfall, i.e., the difference between a reference price, e.g., the current mid-price and the executed trade price. By calculating costs over a range of horizons and sizes, we can build a surface of costs, referred to as a \emph{liquidity risk surface} as a summary of liquidity risk in a market. Also, given its risk aversion/limit or financial obligations, a trader can optimise its strategy and horizon -- longer horizon implies smaller market impact, but exposure to larger market risk while shorter horizon implies larger market impact and smaller market risk.

The structure of the paper is as follows. First, we review related work on agent-based market simulators, transaction cost and market impact models. We then describe our novel ABM for liquidity risk modelling in Sections~\ref{sec:ABM_sim} and \ref{sec:lrc} and present our results in Section~\ref{sec:res}. In Section~\ref{sec:implication}, we discuss the implication of our model for liquidity risk managers and exchanges. Section~\ref{sec:conclusion} concludes and discusses future work.

\section{Related Work}
In this section, we first review the work on agent-based market simulation and on liquidity risk and market impact models and discuss the shortcomings of those models often oversimplified and based on unrealistic assumptions.

\subsection{Agent-based Market Simulators}
A number of different market simulation models have been developed, each with its own strengths and weaknesses. We have a class of models such as the Chiarella model \cite{chiarella2002simulation} that models the demand and supply of different types of trader agents with a simple approximation of price impact and without a limit orderbook. Recent years have seen a class of ABM models that explicitly model the market auctions with limit orderbooks and show to reproduce realistic market behaviours. Some of the popular models are Zero Intelligence Model \cite{farmer2005predictive,preis2006multi,gode1993allocative}, ABIDES \cite{byrd2020abides,vyetrenko2020get,coletta2022learning}, MAXE \cite{belcak2020fast} and the McGroarty's model \cite{mcgroarty}.

\subsection{The Zero-Intelligence (ZI) Model}\label{sec:zi}
Models that seek to reproduce the market dynamics assume that the evolution of the market can be described by a single function, $M_{t+1} = f(M_t)$. An ABM approximates this function by defining explicit rules for the mechanics of the exchange and the behaviour of traders. Hence, the Zero-Intelligence model can be reduced to $f_{\text{zi}}$, where $\mathcal{M}_{t + 1} = f_{\text{zi}}(\alpha_{\text{zi}}, \mu_{\text{zi}}, \delta_{\text{zi}}, \lambda_{zi}, p_{\text{mid}}(t))$, such that a minimally complex trading strategy based on a few parameters can be easily fitted to the observed historical data. For example, the arrival of limit orders can be modelled as a Poisson process with each Zero-Intelligence trader using a Bernoulli variable with parameter $\alpha_{\text{zi}}$ to determine whether it will submit or not a limit order at each step, as are the market orders with a parameter $\mu_{\text{zi}}$ \cite{farmer2005predictive}. The decision to buy or sell is random and not conditioned on the market state and the limit order is submitted at the price level $p_{zi}$  calculated as\footnote{We ignore tick sizing in our mathematical description, but all price levels are at multiple of a tick size defined by the exchange.}:
\begin{align}\label{eq:zi}
    p_{zi}(t) = \left\{
  \begin{array}{@{}ll@{}}
    p_{\text{mid}}(t - 1) - \delta p\text{zi}, & \text{if}\ Buy \\
    p_{\text{mid}}(t - 1) + \delta p\text{zi}, & \text{otherwise}
  \end{array}\right.
\end{align}

\noindent where $\delta p_\text{zi} \sim \exp(\lambda_\text{zi})$ and there is a probability of $\delta_\text{zi}$ to cancel every order on the LOB at each step and $p_{\text{mid}}$ is the mid-price. As \cite{farmer2005predictive} argues, the ZI model can explain certain properties of a financial market despite its simplicity but lacks the complexity for a realistic market simulator.


\subsubsection{ABIDES}
This ABM framework adopts a multi-agent approach with trending, mean-reverting and noise traders. ABIDES is a very flexible model that can be used to simulate a wide variety of market experiments, including for market impact modelling \cite{balch2019evaluate}. The model's $greed$ parameter determines the form of the market impact, but it is unclear how this parameter should be calibrated to calculate a transaction cost in a real market. Furthermore, \cite{balch2019evaluate} looks at market impact of single market orders while \cite{coletta2022learning} tests a Percentage of Volume execution over a short 30-minute execution horizon of a percentage of the average trader volume during the execution period, with no clear indication of the square-root law emerging in either case.

\subsubsection{MAXE}
This ABM framework uses a simple modelling paradigm with zero-intelligence traders \cite{bouchaud2002statistical}, but as with the ABIDES paper \cite{balch2019evaluate}, this work also reports on market impact for single market orders, but not for a meta-order over a horizon.

\subsubsection{The McGroarty model}
This ABM is designed to capture different complex trading behaviours. However, we observe a flaw in the modelling approach where at each step, each agent tries to cancel a single of their orders queued on the LOB. This behaviour does not result in a stable system whereby the depth of the limit orderbook would converge as observed in a real market.

\subsection{Transaction cost and Market Impact Models} \label{sec:mi}
Our work looks at the complex problem of calculating transaction cost whether it is to optimise execution or to understand the liquidity risk associated with a trader's position in the market, which for any number of reasons the trader might need to liquidate at short notice. At the core of liquidity risk management lies a simple question - how does size of a meta-order relate to its market impact. Many models including the seminal Kyle model \cite{Kyle1985} assume a linear impact and models impact as a process by which revealed information is factored into the price.

\begin{figure}
    \centering
    \includegraphics[width=0.8\linewidth]{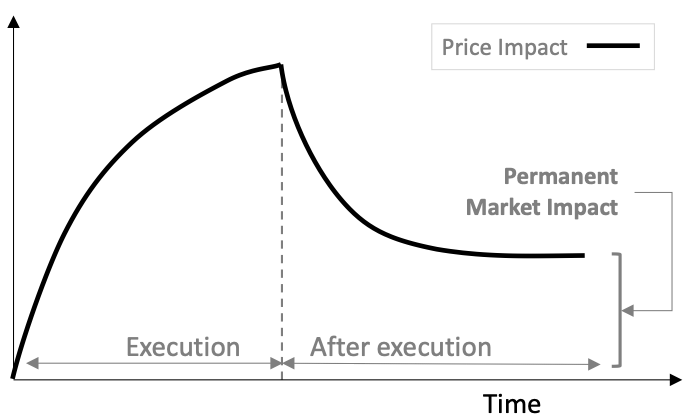}
    \caption{Transient Market Impact where impact temporarily increases, following a square-root law before decaying to a permanent market impact after the last execution}
    \label{fig:ideal_impact}
\end{figure}

Most transaction cost model are simple, assuming a multiplicative power-law relationship of transaction cost and volume to model the cost from market impact and a spread cost (usually defined as crossing the spread from the mid-price) \cite{velu2020algorithmic}. A commonly used transaction cost model is the proprietary Bloomberg Transaction Cost Analysis which is believed to take the following form:
\begin{equation}
    TC = \alpha_B \sigma^{\delta_B} \left(\frac{Q}{ADV}\right)^{\gamma_B} + \beta_B \Delta_{sp} = \frac{1}{3} \sigma \sqrt{\frac{Q}{ADV}} + 0.5 \Delta_{sp}
\end{equation}
\noindent where $Q$ is the volume to execute, $ADV$ is the average daily volume, $\sigma$ is the daily volatility and $\Delta_{sp}$ is the average bid-ask spread. Such models implicitly assume a square-root relationship. The square-root law is observed in the data, but measuring impact requires detailed data and knowledge of hundreds of thousands of proprietary meta-order executions not available in most public dataset \cite{bacry2014market}. Some published studies show the square-root law exists and across different markets, including equities, futures, FX and crypto \cite{bouchaud2022inelastic} and the price market impact takes the form shown in Figure \ref{fig:ideal_impact}.

There are many different transaction cost models \cite{madhavan2000market, bouchaud2009markets,gatheral2010no}, all trying to make sense of market impact that we know exists, is not directly observable and is very difficult to model without oversimplifying or unrealistic assumptions. Next, we describe two of the more popular stylised models regarded as the foundation of market microstructure research \cite{velu2020algorithmic,bouchaudtrades}.

\subsubsection{Kyle Model}
One popular class of equilibrium permanent market impact model is the Kyle Model \cite{Kyle1985}. The model assumes two types of traders in the market - informed traders and uninformed traders. Informed traders have private information about the future value of the asset, while uninformed traders do not. The informed traders trade on their private information by placing orders in the market. As they do so, they reveal some of their information to the uninformed traders and this information leakage causes the price of the asset to move in the direction of the informed trader's order. The model also shows that the informed traders can minimize their market impact by trading slowly and in small increments. The Kyle model however leads to a linear market impact law and fails to capture the square-root law, with no decay of impact.

\subsubsection{Almgren-Chriss Model} \label{sec:almgren}
Their approach approximates the market price formation as a random walk with a model of permanent market impact and a model of transient market impact. Almgren and Chriss' seminal paper \cite{almgren2001optimal} on optimal execution presents a framework to calculate transaction cost of executing a meta-order and to compare different strategies with different time-dependent rates at which the meta-order is executed. The framework can identify the \emph{optimal} strategy given a trader's risk aversion $\lambda$ by assigning a utility to a strategy, measured as the total cost of execution plus a risk aversion towards the variance of the execution cost given a trader's preference. The optimal strategy is then the solution to:
\begin{equation}
    \arg\min_{x}\  E(x) + \lambda V(x)
\end{equation}

\noindent where $E(x)$ is the average execution cost of Strategy $x$ and $V(x)$ is the variance of execution cost. Given a set of strategies, a series of $E(x)$ and $V(x)$ can be calculated, with an efficient frontier of execution calculated as the optimal strategy $\forall \  \lambda \in (0, \infty)$. A strategy that tries to execute the total volume immediately will have maximum cost and a variance of 0 while a balanced, risk-neutral strategy is likely to have the lowest average cost. While a powerful framework, the oversimplification of the problem, e.g., a random walk to approximate the complex market behaviour, means that this class of execution cost model is often not fit for purpose, particularly for shorter horizons and intraday and does not quite capture the complexities of the market microstructure.



\section{The Agent-based Market Simulator}\label{sec:ABM_sim}
Our ABM approach models the mechanics of the exchange and the heterogeneous trading behaviours of the market. The model can be calibrated to a specific day or period. The aim of this model is to capture the statistical properties of the market, while creating a framework upon which more complex behavioural models or complex structures, e.g., to test \emph{a-priori} a change in pricing rules, can be built. It is not one complex model, but a set of simple behavioural models and rules, with model parameters that can be calibrated to reproduce the complexity of the market microstructure. In this section, we describe our ABM in detail.

\subsection{The Structure of the ABM}
The structure of the ABM is effectively a high-frequency discrete-step simulator of a network of \emph{Trader} agents connected to an \emph{Exchange} agent. At each step (of 20 milliseconds for the Hang-Seng Futures Index to average one order per step), \emph{Trader} agents can send messages, e.g., FIX-protocol exchange messages, to submit limit or market orders and amend or cancel queued limit orders and the Exchange sends back execution reports to the Trader as well as Level II market data, i.e., the top ten levels in the buy and sell LOB. 

The Exchange implements a Continuous Double Auction (CDA) \cite{soton263234} which is defined by a set of protocols that specify how the unmatched limit order books are queued in a buy and sell limit order book, how orders are matched, and trades priced, i.e., on price-time priority where the orders are prioritised first by price and then time submitted. The Exchange matching engine will attempt to match each order sequentially. When a sell limit order is less or equal to a buy limit order, a trade occurs and is priced at the earlier order and completely matched orders are removed from the LOB. Unmatched or partially matched limit orders are queued on the LOB. A market order is immediately matched at any price and unmatched market orders are cancelled.

In this paper, we apply our methodology on the Hang-Seng Futures Index market where continuous trading starts at 09:15 and ends at 16:30. In line with existing work in the literature, we restrict our simulation and meta-order execution to continuous trading\footnote{Given our ABM approach, we can model other auction mechanisms of financial markets such as the pre-open call auction before the market opens or pre-close call auction right before the market closes, but it is not within the scope of this paper.}.

\subsection{The Behaviour of the ABM}
Our ABM builds on the Zero-Intelligence model with a data-driven approach, calibrating the trading strategy of the trader to observed historical data, i.e., when and how the trader places a limit or a market order or cancels an order. 

First, we can observe in Section~\ref{sec:zi} that $\alpha_\text{zi}$ and $\mu_\text{zi}$, the rate at which limit and market orders are submitted can be estimated directly from historical rate of arrival of limit and market orders, assuming a Poisson model. To capture the non-uniform intraday liquidity patterns in the market, we estimate during each minute a constant rate of arrival of limit and market orders and scale to the arrival rate at each step $t$ as $\alpha(t)$ and $\mu(t)$ respectively.

Second, we can replace the ZI model's exponential distribution to sample limit order prices with a probabilistic model of depth $\delta p$, volume $v_l$ and duration $\delta t$ of a limit order or and the volume of a market order $q_m$, conditional on the state of the LOB, $\mathcal{M}_{t}^{LOB}$. Within the scope of this paper, we limit $\mathcal{M}_t^{LOB}$ to two important features, i.e., the market spread and time of day. We fit a conditional joint probability distribution $\mathcal{F}_l\left(\delta p, q_l, \delta t | \mathcal{M}_{t}^{LOB}\right)$ to the set of historical limit orders by identifying the depth at which each order was placed, the volume and when the order was cancelled\footnote{Note that with the Zero-Intelligence strategy, cancelling a order with a fixed probability of $\delta$ at each step is equivalent to drawing the duration of the order from an exponential distribution with parameter $\frac{1}{\lambda}$. The latter is computationally more efficient with a single call to a random number generator.} and $\mathcal{F}_m\left(q_m | \mathcal{M}_{t}^{LOB}\right)$ to a set of historical market orders volumes. Conditional on $\mathcal{M}_{t}^{LOB}$, an agent $i$ can sample a limit order or a market order from the joint probability distributions. Then, at each step $t$, an agent $i$ has a probability $\mu(t)$ to submit a market order at volume $q_m$ and a probability $\alpha(t)$ to submit a limit order at sampled volume $q_l$ that is cancelled after $\delta t$ steps and that is submitted at a depth of $\delta p$ and a price level, $p(t)$ calculated as:
\begin{align}
    p(t) = \left\{
  \begin{array}{@{}ll@{}}
    p_{\text{ba}}(t - 1) - \delta p, & \text{if}\ Buy \\
    p_{\text{bb}}(t - 1) + \delta p, & \text{otherwise}
  \end{array}\right.
\end{align}
\noindent where $p_{\text{bb}}$ and $p_{\text{ba}}$ are the best bid price and best ask price respectively at the top of the orderbooks. 

This brings us to the next evolution of the model to capture not only random behaviours, but also alpha behaviours, i.e., what drives a trader to buy or sell. ZI traders effectively assume an efficient market with no information on the direction of the price path and randomly buys or sells. Here, we replace the ZI traders with alpha strategies that decides whether to buy or sell based on the market state. We build on the work on Majewski \emph{et al.} \cite{majewski2020co} to create a new set of traders, \emph{Chiarella} traders  with three different behaviours, i.e., fundamental, momentum and noise traders. Given the demand from a Chiarella trader, $D(t)$, the probability this trader submits a limit order is $\alpha(t) D(t)$ and a market orders is $\mu(t) D(t)$. Next, we summarise the behaviour of the different Chiarella Traders with different demands.

\subsubsection{Fundamental Traders}\label{sec:FT}
Given a random walk model of the fair and exogeneous fundamental value of the traded instrument, $V_t$, the demand of fundamental traders, $D^f(t)$ is a function of the distortion between current market price and $V_t$. We introduce a new concept, the \emph{reflexive fundamental value} $\tilde{V}_t$. On top of the exogeneous signal $V_t$, we also account for entropy within a simulation. As individual trades are executed, they reveal information about the true underlying demand and supply, that the fundamental trader factors in when estimating the fair value of the traded instrument:
\begin{align}\label{fig_fun}
    D^f(t) = \left\{
  \begin{array}{@{}ll@{}}
    \kappa \left(\tilde{V}_t - p_{\text{ba}}(t)\right), & \text{if}\ V_t > p_{\text{ba}}, \\
    \kappa \left(\tilde{V}_t - p_{\text{bb}}(t)\right), & \text{if}\ V_t < p_{\text{bb}}, \\
    0 & \text{otherwise}
  \end{array}\right.
\end{align}
\noindent where,
\begin{align}
    V_t &= V_{t-1} + g_V \  dt + \sigma_V \  dW_t^{(V)} \\
    \tilde{V}_t &= V_t + X_t \\
    X_t &= X_{t-1} + f_\text{mi}(Q_t)
\end{align}
\noindent where $Q_t$ is the excess demand\footnote{At each step, we aggregate trade volume as $Q_t$ and if traded price $>$ mid-price, we assume an excess demand $Q_t$, else, an excess supply, $-Q_t$.} at step $t$. In Section \ref{sec:lrc}, we will demonstrate how permanent market impact emerges from this reflexive fundamental value. Furthermore, we note that $g_V$ can be set depending on the modeller's belief of whether the market will go up or down. For our liquidity risk management use case, we assume that there is no drift in our forward path simulation, i.e., $g_V = 0$ and the past cannot predict future price movement. We refer the reader to a detailed discussion in \cite{almgren2001optimal} of a drift when calculating liquidity risk.

\subsubsection{Momentum Traders}
Momentum traders buy on an up trend and sell on a down trend of market mid-prices, $p_\text{mid}(t)$, contributing to trending patterns in the market. We adopt two types of momentum traders each with a set of $\beta_H$ and $\gamma_H$ for high-frequency momentum traders with an arbitrarily high $\eta_H$ and a set of $\beta_L$ and $\gamma_L$ for for low-frequency momentum traders with an arbitrarily low $\eta_L$. The demand of a momentum trader $D^{M}(t)$ is defined as:
\begin{align}
    D^{M}(t) &= \beta \tanh\left(\gamma M(t)\right) \\
    M(t) &= (1 - \eta) \  M(t-1) dt + \eta \left(p_\text{mid}(t) - p_\text{mid}(t-1)\right)
\end{align}

\subsubsection{Noise Traders}
Noise traders randomly buy or sell and their demand $D^N(t)$ effectively follow a random walk:
\begin{align}
    D^N(t) &= \sigma \zeta, \text{where }\zeta~\sim~N(0,1)
\end{align}
With these more complex behaviours, the arrival of orders resembles a Hawkes Process rather than a Poisson Process as observed in a Zero-Intelligence model, capturing long-range persistence of orderflow (buy more likely to be followed by a buy order and a sell by a sell order) \cite{bouchaudtrades}. Against this overview of the different trading behaviours, we next look at calibrating our ABM.


\subsection{Calibrating to Historical Market Data}\label{Sec:ABM_cal}
\begin{figure}[h]
    \centering
    \includegraphics[width=1\linewidth]{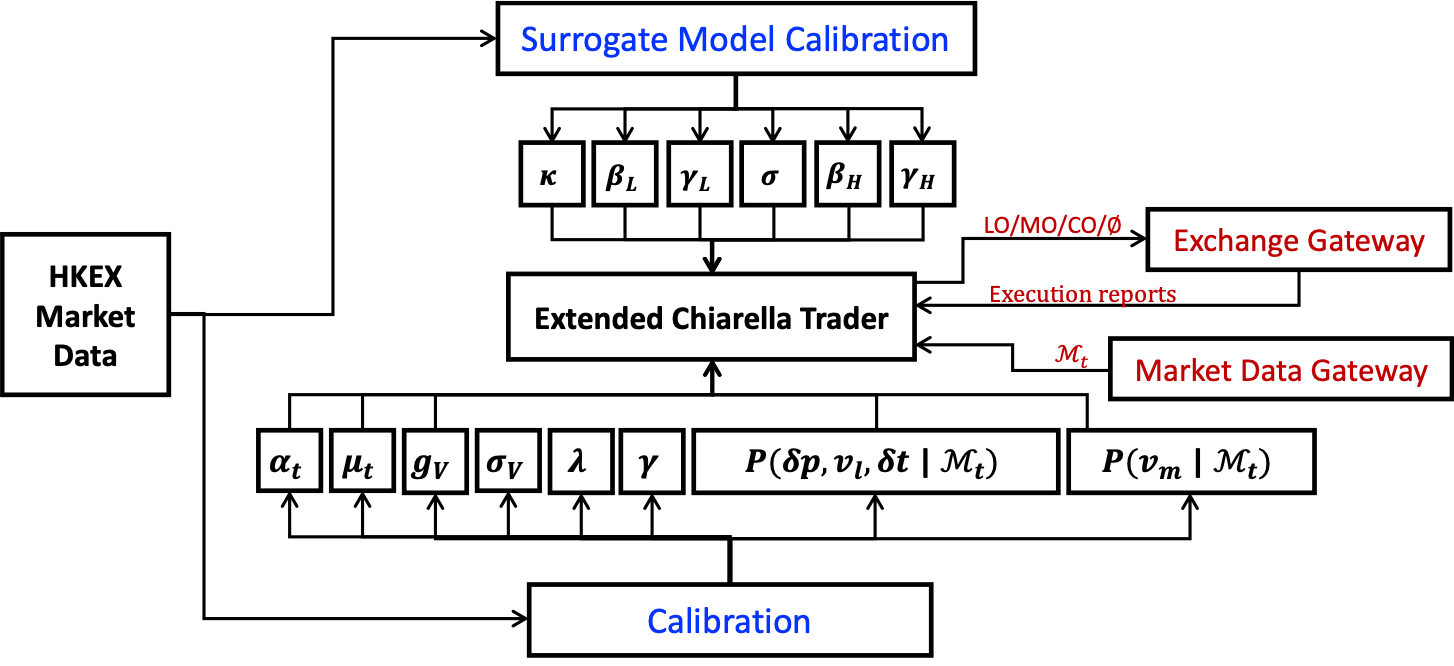}
    \caption{The ABM market simulator dependencies on market data and model parameters}
    \label{fig:eib}
\end{figure}

The ABM dependencies are summarised in Figure \ref{fig:eib}. 

\subsubsection{The Market Tick and Limit Orderbook Data}
The ABM is calibrated to end-of-day Hang-Seng Index Futures tick dataset on 2022-12-23\footnote{The most liquid front contract, HSIZ2, was used.} which consists of 3.4M new, amend, cancel limit order operations and 85k trades. By rebuilding the limit orderbook from the sequence of operations, at each tick, we can build the L2 market data, i.e., the price and volume at the top ten levels of the orderbook as well as a list of all limit orders submitted and finally, we can infer from trades a list of all market orders.


\subsubsection{Calibrating to observed data}
Parameters $\alpha(t)$, $\mu(t)$ are calibrated to the rate of arrival of limit and marker orders every minute, assuming constant rate at each step $t$ during each minute. Given that we introduce a reflexive fundamental value that will add in the impact from individual trade at each step, we remove the cummulative market impact from historical trade prices $\hat{V}_t - \sum_{i=0}^{t}f_\text{mi}(Q_i)$ to get a reasonable a proxy of the exogeneous fundamental value $V_t$. Then, $\sigma_V$ is calibrated to the proxy signal under the assumption of a random walk with no drift. 

\begin{figure*}[h]
    \centering
    \includegraphics[width=1\linewidth]{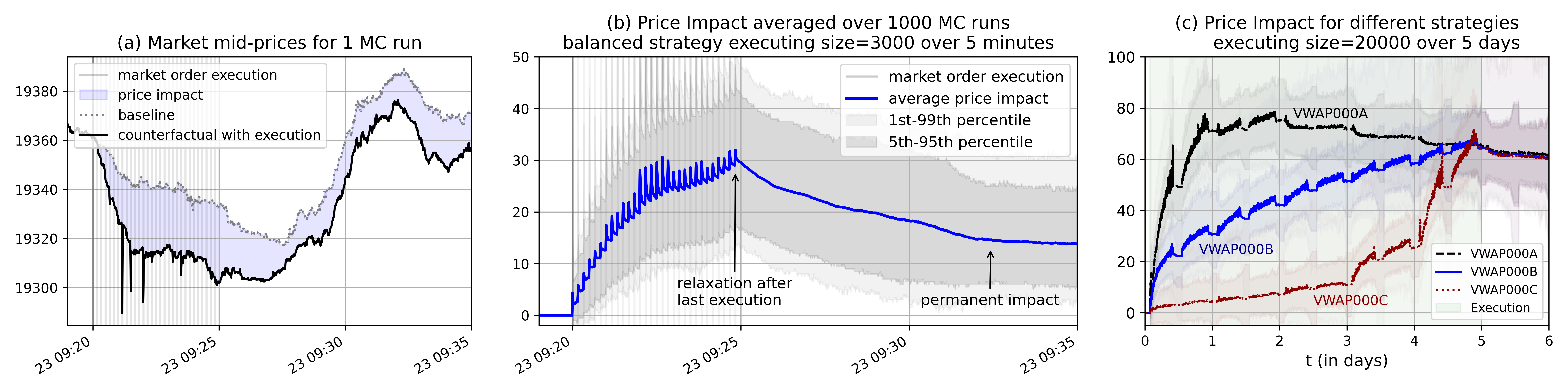}
    \caption{Transient and permanent impact emerging from simulating a baseline and a counterfactual in (a) and (b). In (c), we empirically demonstrate the relevance of the execution strategy on price impact. `VWAP000A' is a front-loaded strategy, executing 70\%, 20\%, 5\%, 3\% and 2\% on each day. `VWAP000B' is balanced and 'VWAP000C' is a back-loaded strategy with the reverse schedule of `VWAP000A'.}
    \label{fig:mi2}
\end{figure*}

\begin{figure*}[h]
    \centering
    \includegraphics[width=1\linewidth]{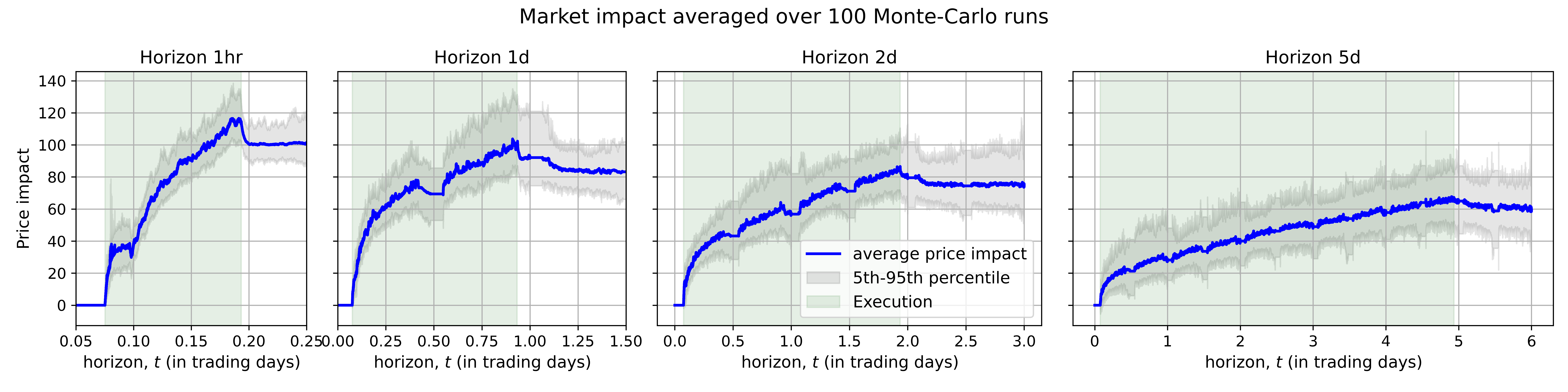}
    \caption{Price impact when executing a size of 20,000 over different horizons.}
    \label{fig:mi3}
\end{figure*}

Next, we empirically estimate the conditional joint probability function $\mathcal{F}_l(\delta p, q_{L}, \delta t | \mathcal{M}_{t}^{LOB})$ by identifying the time at which each historical order was submitted, the spread before they were submitted, the size of the order, the duration in steps and the depth at which they we placed. Then, conditional on the current spread and time in the simulation, we estimate a joint distribution over price, volume and duration. To avoid modelling the correlation structure of the joint distribution, we adopt an empirical approach where we use historical simulation and draw a random sample from a subset of all historical orders conditioned on spread and time to get a depth, volume and duration whenever a Chiarella Trader submits a limit order. Similarly,  $\mathcal{F}_m(q_M | \mathcal{M}_{t}^{LOB})$ is calibrated to historical market order volumes. 

\subsubsection{Calibrating the single-trade impact model}
Next, we calibrate the market impact of an individual trade as an aggregate impact function:
\begin{equation}\label{eqn:market_impact}
    f_\text{mi}(Q) = \lambda Q^{\gamma}
\end{equation}

There are different approaches to modelling price impact of a single trade \cite{bouchaudtrades}. We adopt the empirically most robust approach for a high-liquidity derivative as the Hang-Seng Futures Index, i.e., the aggregate impact function. Specifically, we calculate impact over a time scale rather than trade-by-trade. We aggregate at each second the orderflow imbalance, i.e., the positive buy trade volumes and negative sell trade volumes as excess demand and calculate the mid-price change as the price impact during that period. Under the assumption that a positive impact for excess demand and a negative impact for excess supply will emerge as market risk cancels out on average, we fit $f_\text{mi}$ to the observed excess demand and impact data using least-square regression. We note that for HSIZ2 trade data on 2022-12-23, the calibrated model $f_\text{mi}(Q) = 0.561 \sqrt{Q}$ shows similar concavity to the results reported by Bouchaud \emph{et. al.} \cite{bouchaudtrades}.


\subsubsection{Calibrating the Chiarella Model Parameters using surrogate modelling}
While we can observe parameters such as arrival of limit orders, we do not know the trading behaviour behind each order and, so, the hidden Chiarella model parameters $\kappa$, $\beta_L$, $\beta_H$, $\gamma_L$, $\gamma_H$, $\sigma$ need to be estimated as the parameters that most accurately reproduces the market microstructure. This is an optimisation problem where we minimise the distance between historical market and the simulated market. We quantify the difference between two markets by approximating them to a set of measurable stylised facts and the distance to minimise is the sum of distances between the different stylised facts \cite{kang2022} given a set of model parameters. The function to minimise is stochastic in nature and we solve the optimisation using a well-known approach, surrogate modelling \cite{lamperti2018agent} that reduces the number of required simulations and is well suited for calibrating the market simulation as each function call requires running a multi-agent simulation with many agents and millions of steps that can take seconds to minutes. 
We optimise the Chiarella model parameters\footnote{For HSI on 2022-12-23, we set $\eta_H = 0.98$ and 
$\eta_L = 1.7e^{-4}$ and calibrated $\kappa = 0.011$, $\beta_L  = 1.976$, $\gamma_L = 5.26$, $\beta_H = 0.530$, $\gamma_H = 290000$ and $\sigma = 0.249$.} against the following stylised facts\footnote{We select a subset of stylised facts that best capture the patterns of mid to long-term market returns as we assume the calibrated behaviour to observed data captures the short-term trading behaviour.}:
\begin{itemize}
    \item Arrival rate of limit and market orders at every minute
    \item Distribution of spread
    \item Distribution of 1s, 60s returns and absolute returns
    \item Autocorrelation of 1s, 60s returns and absolute returns
    \item The persistence of orderflow, i.e., the auto-correlation of buy/sell signals
\end{itemize}

\section{The Liquidity Risk Calculator} \label{sec:lrc}
Liquidity risk considers the uncertainties in executing an order in the market and the variable costs to do so attributed to market slippage caused by price impact and market risk.
Market impact is when the market moves against the trader as they execute their meta-order. Market risk includes the price fluctuations under external economic drivers, whether or not the meta-order is executed. With an ABM approach, it is possible to explicitly attribute the cost of execution to either market risk or market impact. In this section, we first introduce this methodology and second, we demonstrate how market impact emerges from the market simulation.

\begin{figure*}[ht]
    \centering
    \includegraphics[width=1.0\linewidth]{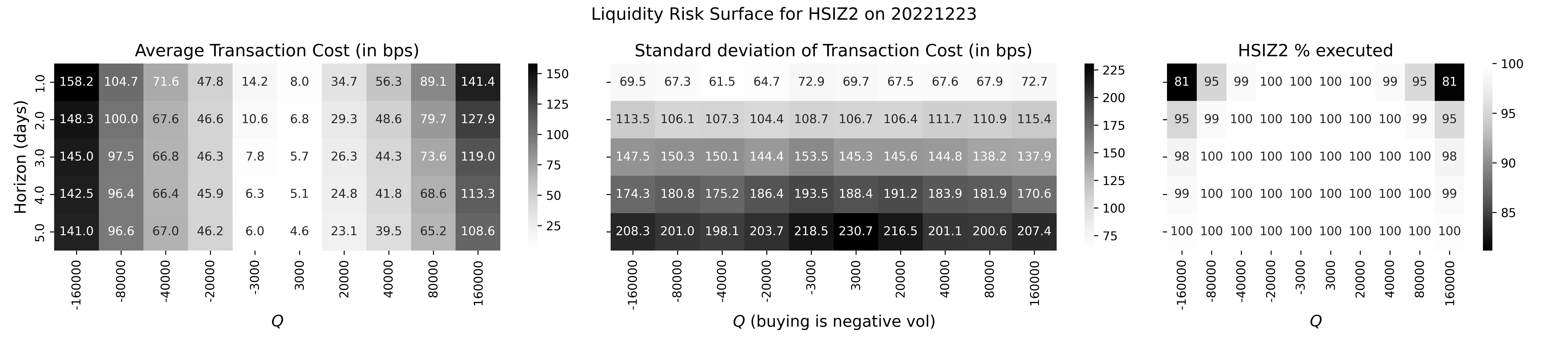}
    \caption{Liquidity risk surface for HSIZ2 calibrated to 2022-12-23. Left panel is the average cost (over costs from market impact), middle panel is standard deviation of cost (over costs from market slippage, i.e., impact and risk) and right panel is \% of schedule executed. Where < 100\% is executed, we extrapolate the surface using a cubic-spline.}
    \label{fig:hsi_20221223}
\end{figure*}

\begin{figure*}[ht]
    \centering
    \includegraphics[width=1.\linewidth]{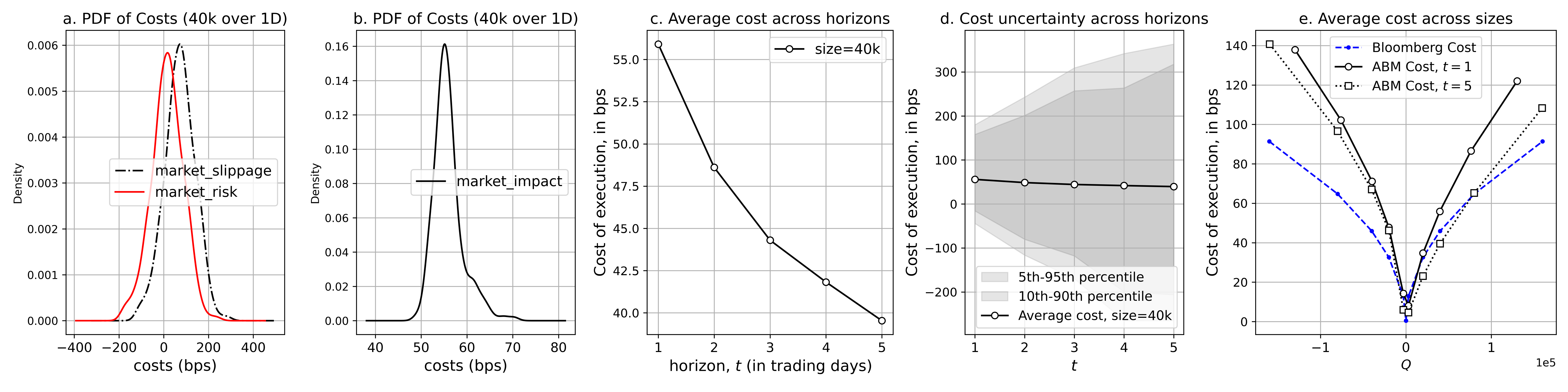}
    \caption{Liquidity risk for HSIZ2 calibrated to 2022-12-23. At a horizon of 0 days, we expect the uncertainty to be very low and equivalent to the standard deviation of the depth of the LOB at $t_0$. Note the low signal-to-noise ratio of market risk costs compared to market impact costs in Panel (a) and (b).}
    \label{fig:3dmi}
\end{figure*}

A novelty of our ABM approach is to decompose transaction cost and attribute uncertainties to market risk and market impact. The cost of execution $\zeta$ is calculated as an implementation shortfall\footnote{The  Implementation Shortfall cost of often given in basis  points, as  $10^4 \times \frac{\zeta}{p_R}$. The reference price is usually the current observed mid-price and we would trade at this price plus a half-spread assuming unlimited liquidity on the orderbook.}, given a reference price $p_R$, 
\begin{align}
    \zeta &= \frac{\sum_{t} p_t v_t^E}{\sum_{t} v_t^E} - p_R
\end{align}  
\noindent where $p_t$ are the executed trade prices and $v_t^E$ are the executed volumes at each step.

We observe that by running a baseline and a counterfactual experiment where the only difference is the execution of the large order given some execution strategy in the counterfactual, using the same random number generator, with each trader using the same random number generator and using the same exogeneous signal $V_t$, we can decompose the transaction cost to cost from market impact $\zeta_{MI}$ and cost from market risk $\zeta_{MR}$:
\begin{align}
    \zeta &= \zeta_{MR} + \zeta_{MI} \\
    \zeta_{MR} &= \frac{\sum_{t} p_t^{B} v_t^E}{\sum_{t} v_t^E} - p_R \\
    \zeta_{MI} &= \frac{\sum_{t} (p_t - p_t^{B})\  v_t^E}{\sum_{t} v_t^E}
\end{align}
\noindent where $p_t^{B}$ are the simulated prices in the baseline experiment with the execution. $\zeta$ can be broken down to $\zeta_{MR}$, the cost with zero market impact and with only exposure to market risk and to $\zeta_{MI}$, the cost due to market impact and not market risk exposure. Market price impact is calculated as the difference in mid-price between the baseline and the counterfactual, $p_{MI} = p_t - p_t^{B}$.

\section{Results} \label{sec:res}
In this section, we present our main results of the cost of execution on the Hang-Seng Futures Index. Specifically, we first demonstrate how our price impact emerges in our ABM and we then build a \emph{liquidity risk surface} of average cost and uncertainties for different execution horizons and sizes for the Hang-Seng Futures Index.

\subsection{Emerging market impact}
In our first experiment in Figure {fig:mi2}(a) and (b), we liquidate a meta-order of 3000 into multiple market orders executed every 10s over 5 minutes. We do not model market impact of this meta-order in our model, but instead, we observe a negative transient and permanent price impact emerge from the simulation and follows the square-root law described in the literature, see Section~\ref{sec:mi}. From Figure \ref{fig:mi2}(c) and Figure \ref{fig:mi3}, we also note that impact depends on the execution horizon and the execution strategy and a framework that calculates price impact of executing a meta-order of any size, over any horizon whether seconds or days and given any strategy with a combination of limit and market orders is very promising for the industry and can address the limitations of stylised models. Future work will analyse the relationships between impact and size, horizon and strategy in more detail.



\subsection{The Hang-Seng Index Liquidity Risk}\label{sec:lrs}
Under realistic impact, we now consider the cost of execution for HSI Futures Index (front month HSIZ2 contract) on 2022-12-23. For each configuration over a grid of horizons and sizes, we calculate the transaction costs for 400 Monte-Carlo runs. Because we assume no drift (see Section~\ref{sec:FT}), we expect the average cost from market risk to converge to zero such that the average cost of execution converges to the average cost of execution from market impact which converges faster. Given the significantly lower signal to noise ratio of market risk cost, we calculate a more accurate estimate of average cost from market impact. Note that when drift is factored in, the average cost should be from both market risk and market impact with a non-zero average market risk cost as the market drifts. The standard deviation captures uncertainties from both market risk exposure and market impact.

Intuitively, as we increase $Q$, we expect costs to increase and as we increase the horizon, we expect less impact on the market and costs to decrease. The results in Figure \ref{fig:hsi_20221223} are in line with our expectation. From Figure \ref{fig:3dmi}(c) and (d), we observe a gradual decay of costs and an increase in uncertainty of costs as we increase the horizon. We also observe in Figure \ref{fig:3dmi}(e) a concave increase in costs as we increase $Q$, a square-root law often  described in the literature \cite{bacry2014market}. We compare our results to the Bloomberg Transaction Cost model\footnote{We calibrate $\sigma$ of the Bloomberg Transaction Cost model to a history of three months to $432.7$ and return the costs in basis points (bps).}, a function of executed size and not horizon (see Section~\ref{sec:mi}), in Figure \ref{fig:3dmi}(e) which, we observe, underestimate the liquidity risk compared to our model. However, in practice, $\alpha_B$, $\delta_B$, $\gamma_B$ and $\beta_B$ would be calibrated to HSI based on proprietary execution data and as a sanity check, we are satisfied our ABM model produces transaction costs similar in magnitude to the commonly used approximation of the Bloomberg Transaction Cost model.

\subsection{Optimal Execution}\label{sec:oe}
The transaction cost also depends on the execution strategy, whether on how to slice and schedule smaller orders or how to execute each order. Our ABM framework can compare and optimise strategies. Within the scope of this paper, we use the Almgren-Chriss framework for optimal execution - see Section~\ref{sec:almgren} with a simple set of strategies that define the volume scheduled for each day over five days and a VWAP strategy\footnote{In our VWAP strategy, the meta-order sliced into market orders scheduled every 5s and volumes are set to execute larger volumes at times when the market is historically more liquid intraday.}.

\begin{figure}[h]
    \centering
    \includegraphics[width=1\linewidth]{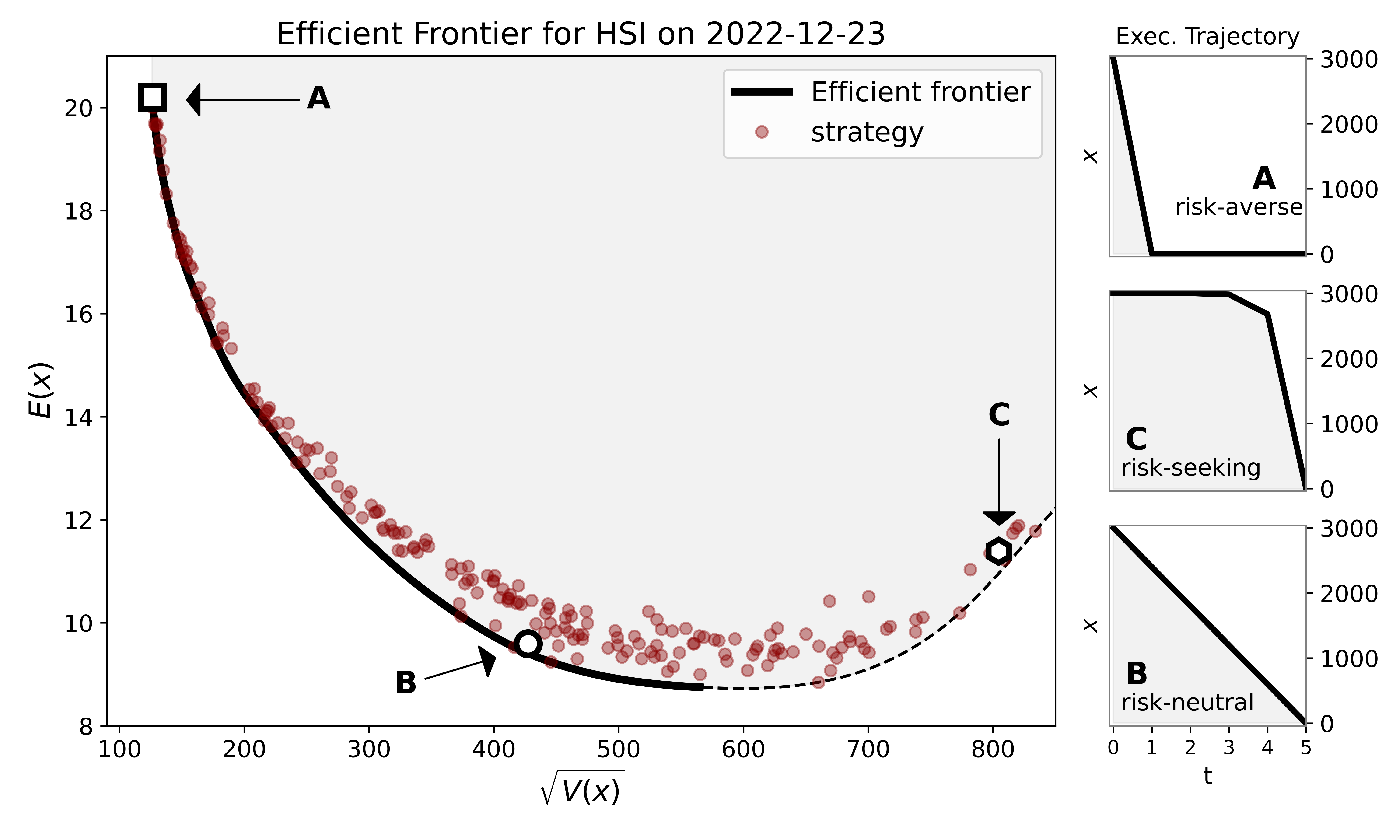}
    \caption{The ABM Efficient Frontier for optimal execution with 200 sampled strategies. Strategy A liquidates the total volume on Day 1 - variance is not 0 due to some risk exposure intraday. Balanced strategy B executes 20\% daily for 5 days. Strategy C executes at the end and is always sub-optimal under the no-drift assumption. The dashed curve represents sub-optimal strategies with higher uncertainty for the same average costs. $x$ is the remaining volume to execute at time $t$.}
    \label{fig:ef}
\end{figure}

Given different strategies to execute a meta-order of 3000 over five days, our ABM model is able to reproduce similar results as the Almgren-Chriss framework -- see Figure \ref{fig:ef} and \cite{almgren2001optimal}. Specifically, we observe the front-loaded strategy $A$ has the lowest uncertainty and highest cost as expected. Balanced strategy $B$ is close to the minimum and with enough runs, we expect balanced strategy $B$ to converge that minimum. Overall, our ABM framework allows us to capture the complexities of strategies and a trader can use this framework to test out its strategies or even optimise parameters of an execution algo to minimise its risk-adjusted cost.

\section{Implications for Exchanges and Liquidity Risk Managers} \label{sec:implication}
Our ABM liquidity risk model has interesting implications for an exchange \cite{hkex}. Specifically, they can run simulation to assess the impact of a participant defaulting and manage close-out activities in a clearing exchange. For a Central Counterparty Clearing House (CCP), when a participant is unable to fulfil its liability after margin call, the CCP will take actions to liquidate the default participant positions. The forced sales drive prices down and margin calls act as feedback to magnify the effects, triggering more selling. An important challenge is to balance the liquidity risk of default management, market stability after the fire sale and mitigate systemic risk. So, the CCP needs to identify the optimal liquidation strategy within management risk limits. Our ABM model provides such a framework as demonstrated in Sections~\ref{sec:lrs} and \ref{sec:oe} for default management and to better understand liquidity risk in the market.

Similarly, liquidity risk managers are able to build a more accurate view of their liquidity risk specific to their liquidation policies and to understand the liquidity risk associated with large positions in the market. Our ABM model can also be used to optimise their liquidation strategies, whether to decide on the liquidation horizon or the execution strategy.

\section{Conclusion and Future Work}\label{sec:conclusion}
In this paper, we introduce a novel agent-based high-frequency market simulation that is calibrated to historical tick data and demonstrate how such a model can be used for liquidity risk management, whether it is to calculate the cost of executing a meta-order or to optimise a strategy. We discuss how market impact is hard to model or even observe in the real world as it impossible to know what could have been if the meta-order had not been executed. In our ABM, we do not explicitly model either transient or permanent impact, but our model is based on intuitive trader behaviours, on a model of single-trade price impact and the reflexive fundamental value, a new signal that drive fundamental traders that assume orderflow is informed and reveals information, besides the exogeneous fundamental value. The result of this novel approach is realistic transient and permanent market impact that emerge, that are in-line with what practitioners would expect from the market, e.g., follows the square-root law and relaxes to a permanent level. We discuss how our ABM framework can be a valuable tool for real-world practitioners such as exchanges and liquidity risk managers, for example to better understand the risk profile of large positions in the market.

In the future, we intend to improve on the execution strategy of the Chiarella Traders to use the whole L2 market data (rather than only two features, i.e., time and spread) to drive their execution logic, e.g., we could train a deep neural network to estimate the joint distribution of order price, volume and duration given several features e.g., the top ten levels of the LOB, order flow imbalance, momentum at different scales and volatility at different scales and whether they are buying or selling. Their alpha decision to buy or sell would be driven by the Chiarella strategy. Furthermore, we could investigate other model of single-trade market impact that would be robust even in low-liquidity market with few data points. Finally, we could extend our model to multiple symbols that accurately capture the correlation structure of market returns and orderbook dynamics and to emerge cross market impact, i.e., how executing a large order in a market can affect another.







\begin{thebibliography}{31}


\ifx \showCODEN    \undefined \def \showCODEN     #1{\unskip}     \fi
\ifx \showDOI      \undefined \def \showDOI       #1{#1}\fi
\ifx \showISBNx    \undefined \def \showISBNx     #1{\unskip}     \fi
\ifx \showISBNxiii \undefined \def \showISBNxiii  #1{\unskip}     \fi
\ifx \showISSN     \undefined \def \showISSN      #1{\unskip}     \fi
\ifx \showLCCN     \undefined \def \showLCCN      #1{\unskip}     \fi
\ifx \shownote     \undefined \def \shownote      #1{#1}          \fi
\ifx \showarticletitle \undefined \def \showarticletitle #1{#1}   \fi
\ifx \showURL      \undefined \def \showURL       {\relax}        \fi
\providecommand\bibfield[2]{#2}
\providecommand\bibinfo[2]{#2}
\providecommand\natexlab[1]{#1}
\providecommand\showeprint[2][]{arXiv:#2}

\bibitem[Almgren and Chriss(2001)]%
        {almgren2001optimal}
\bibfield{author}{\bibinfo{person}{Robert Almgren} {and} \bibinfo{person}{Neil Chriss}.} \bibinfo{year}{2001}\natexlab{}.
\newblock \showarticletitle{Optimal execution of portfolio transactions}.
\newblock \bibinfo{journal}{\emph{Journal of Risk}}  \bibinfo{volume}{3} (\bibinfo{year}{2001}), \bibinfo{pages}{5--40}.
\newblock


\bibitem[Bacry et~al\mbox{.}(2014)]%
        {bacry2014market}
\bibfield{author}{\bibinfo{person}{Emmanuel Bacry}, \bibinfo{person}{Adrian Iuga}, \bibinfo{person}{Matthieu Lasnier}, {and} \bibinfo{person}{Charles-Albert Lehalle}.} \bibinfo{year}{2014}\natexlab{}.
\newblock \showarticletitle{Market impacts and the life cycle of investors orders}.
\newblock  (\bibinfo{year}{2014}).
\newblock


\bibitem[Balch et~al\mbox{.}(2019)]%
        {balch2019evaluate}
\bibfield{author}{\bibinfo{person}{Tucker~Hybinette Balch}, \bibinfo{person}{Mahmoud Mahfouz}, \bibinfo{person}{Joshua Lockhart}, \bibinfo{person}{Maria Hybinette}, {and} \bibinfo{person}{David Byrd}.} \bibinfo{year}{2019}\natexlab{}.
\newblock \showarticletitle{How to evaluate trading strategies: Single agent market replay or multiple agent interactive simulation?}
\newblock \bibinfo{journal}{\emph{arXiv preprint arXiv:1906.12010}} (\bibinfo{year}{2019}).
\newblock


\bibitem[Belcak et~al\mbox{.}(2020)]%
        {belcak2020fast}
\bibfield{author}{\bibinfo{person}{Peter Belcak}, \bibinfo{person}{Jan-Peter Calliess}, \bibinfo{person}{Stefan Zohren}, {et~al\mbox{.}}} \bibinfo{year}{2020}\natexlab{}.
\newblock \showarticletitle{Fast agent-based simulation framework of limit order books with applications to pro-rata markets and the study of latency effects}.
\newblock \bibinfo{journal}{\emph{arXiv preprint arXiv:2008.07871}} (\bibinfo{year}{2020}).
\newblock


\bibitem[Bookstaber et~al\mbox{.}(2018)]%
        {bookstaber2018agent}
\bibfield{author}{\bibinfo{person}{Richard Bookstaber}, \bibinfo{person}{Mark Paddrik}, {and} \bibinfo{person}{Brian Tivnan}.} \bibinfo{year}{2018}\natexlab{}.
\newblock \showarticletitle{An agent-based model for financial vulnerability}.
\newblock \bibinfo{journal}{\emph{Journal of Economic Interaction and Coordination}} \bibinfo{volume}{13}, \bibinfo{number}{2} (\bibinfo{year}{2018}), \bibinfo{pages}{433--466}.
\newblock


\bibitem[Bouchaud et~al\mbox{.}({[n.\,d.]})]%
        {bouchaudtrades}
\bibfield{author}{\bibinfo{person}{JP Bouchaud}, \bibinfo{person}{J Bonart}, \bibinfo{person}{J Donier}, {and} \bibinfo{person}{M Gould}.} \bibinfo{year}{[n.\,d.]}\natexlab{}.
\newblock \bibinfo{title}{Trades, quotes and prices: financial markets under the microscope, 2018}.
\newblock
\newblock


\bibitem[Bouchaud(2022)]%
        {bouchaud2022inelastic}
\bibfield{author}{\bibinfo{person}{Jean-Philippe Bouchaud}.} \bibinfo{year}{2022}\natexlab{}.
\newblock \showarticletitle{The inelastic market hypothesis: a microstructural interpretation}.
\newblock \bibinfo{journal}{\emph{Quantitative Finance}} \bibinfo{volume}{22}, \bibinfo{number}{10} (\bibinfo{year}{2022}), \bibinfo{pages}{1785--1795}.
\newblock


\bibitem[Bouchaud et~al\mbox{.}(2009)]%
        {bouchaud2009markets}
\bibfield{author}{\bibinfo{person}{Jean-Philippe Bouchaud}, \bibinfo{person}{J~Doyne Farmer}, {and} \bibinfo{person}{Fabrizio Lillo}.} \bibinfo{year}{2009}\natexlab{}.
\newblock \showarticletitle{How markets slowly digest changes in supply and demand}.
\newblock In \bibinfo{booktitle}{\emph{Handbook of financial markets: dynamics and evolution}}. \bibinfo{publisher}{Elsevier}, \bibinfo{pages}{57--160}.
\newblock


\bibitem[Bouchaud et~al\mbox{.}(2002)]%
        {bouchaud2002statistical}
\bibfield{author}{\bibinfo{person}{Jean-Philippe Bouchaud}, \bibinfo{person}{Marc M{\'e}zard}, {and} \bibinfo{person}{Marc Potters}.} \bibinfo{year}{2002}\natexlab{}.
\newblock \showarticletitle{Statistical properties of stock order books: empirical results and models}.
\newblock \bibinfo{journal}{\emph{Quantitative finance}} \bibinfo{volume}{2}, \bibinfo{number}{4} (\bibinfo{year}{2002}), \bibinfo{pages}{251}.
\newblock


\bibitem[Byrd et~al\mbox{.}(2020)]%
        {byrd2020abides}
\bibfield{author}{\bibinfo{person}{David Byrd}, \bibinfo{person}{Maria Hybinette}, {and} \bibinfo{person}{Tucker~Hybinette Balch}.} \bibinfo{year}{2020}\natexlab{}.
\newblock \showarticletitle{ABIDES: Towards high-fidelity multi-agent market simulation}. In \bibinfo{booktitle}{\emph{Proceedings of the 2020 ACM SIGSIM Conference on Principles of Advanced Discrete Simulation}}. \bibinfo{pages}{11--22}.
\newblock


\bibitem[Chiarella and Iori(2002)]%
        {chiarella2002simulation}
\bibfield{author}{\bibinfo{person}{Carl Chiarella} {and} \bibinfo{person}{Giulia Iori}.} \bibinfo{year}{2002}\natexlab{}.
\newblock \showarticletitle{A simulation analysis of the microstructure of double auction markets}.
\newblock \bibinfo{journal}{\emph{Quantitative finance}} \bibinfo{volume}{2}, \bibinfo{number}{5} (\bibinfo{year}{2002}), \bibinfo{pages}{346}.
\newblock


\bibitem[Cliff and Bruten(1997)]%
        {cliff1997zero}
\bibfield{author}{\bibinfo{person}{Dave Cliff} {and} \bibinfo{person}{Janet Bruten}.} \bibinfo{year}{1997}\natexlab{}.
\newblock \showarticletitle{Zero is Not Enough: On The Lower Limit of Agent Intelligence For Continuous Double Auction Markets}.
\newblock \bibinfo{journal}{\emph{Technical Report HPL-97-141, Hewlett-Packard Laboratories, Bristol, UK}} (\bibinfo{year}{1997}).
\newblock


\bibitem[Coletta et~al\mbox{.}(2022)]%
        {coletta2022learning}
\bibfield{author}{\bibinfo{person}{Andrea Coletta}, \bibinfo{person}{Aymeric Moulin}, \bibinfo{person}{Svitlana Vyetrenko}, {and} \bibinfo{person}{Tucker Balch}.} \bibinfo{year}{2022}\natexlab{}.
\newblock \showarticletitle{Learning to simulate realistic limit order book markets from data as a World Agent}. In \bibinfo{booktitle}{\emph{Proceedings of the Third ACM International Conference on AI in Finance}}. \bibinfo{pages}{428--436}.
\newblock


\bibitem[Cont and Mueller(2019)]%
        {Cont2019}
\bibfield{author}{\bibinfo{person}{Rama Cont} {and} \bibinfo{person}{Marvin~S. Mueller}.} \bibinfo{year}{2019}\natexlab{}.
\newblock \showarticletitle{A stochastic partial differential equation model for limit order book dynamics}.
\newblock  (\bibinfo{date}{4} \bibinfo{year}{2019}).
\newblock
\urldef\tempurl%
\url{http://arxiv.org/abs/1904.03058}
\showURL{%
\tempurl}


\bibitem[Farmer et~al\mbox{.}(2005)]%
        {farmer2005predictive}
\bibfield{author}{\bibinfo{person}{J~Doyne Farmer}, \bibinfo{person}{Paolo Patelli}, {and} \bibinfo{person}{Ilija~I Zovko}.} \bibinfo{year}{2005}\natexlab{}.
\newblock \showarticletitle{The predictive power of zero intelligence in financial markets}.
\newblock \bibinfo{journal}{\emph{Proceedings of the National Academy of Sciences}} \bibinfo{volume}{102}, \bibinfo{number}{6} (\bibinfo{year}{2005}), \bibinfo{pages}{2254--2259}.
\newblock


\bibitem[Gao et~al\mbox{.}(2022)]%
        {kang2022}
\bibfield{author}{\bibinfo{person}{Kang Gao}, \bibinfo{person}{Perukrishnen Vytelingum}, \bibinfo{person}{Stephen Weston}, \bibinfo{person}{Wayne Luk}, {and} \bibinfo{person}{Ce Guo}.} \bibinfo{year}{2022}\natexlab{}.
\newblock \showarticletitle{{Understanding intra-day price formation process by agent-based financial market simulation: calibrating the extended chiarella model}}.
\newblock  (\bibinfo{date}{Aug.} \bibinfo{year}{2022}).
\newblock


\bibitem[Gatheral(2010)]%
        {gatheral2010no}
\bibfield{author}{\bibinfo{person}{Jim Gatheral}.} \bibinfo{year}{2010}\natexlab{}.
\newblock \showarticletitle{No-dynamic-arbitrage and market impact}.
\newblock \bibinfo{journal}{\emph{Quantitative finance}} \bibinfo{volume}{10}, \bibinfo{number}{7} (\bibinfo{year}{2010}), \bibinfo{pages}{749--759}.
\newblock


\bibitem[Gode and Sunder(1993)]%
        {gode1993allocative}
\bibfield{author}{\bibinfo{person}{Dhananjay~K Gode} {and} \bibinfo{person}{Shyam Sunder}.} \bibinfo{year}{1993}\natexlab{}.
\newblock \showarticletitle{Allocative efficiency of markets with zero-intelligence traders: Market as a partial substitute for individual rationality}.
\newblock \bibinfo{journal}{\emph{Journal of political economy}} \bibinfo{volume}{101}, \bibinfo{number}{1} (\bibinfo{year}{1993}), \bibinfo{pages}{119--137}.
\newblock


\bibitem[Kyle(1985)]%
        {Kyle1985}
\bibfield{author}{\bibinfo{person}{Albert~S. Kyle}.} \bibinfo{year}{1985}\natexlab{}.
\newblock \showarticletitle{Continuous Auctions and Insider Trading}.
\newblock \bibinfo{journal}{\emph{Econometrica}}  \bibinfo{volume}{53} (\bibinfo{date}{11} \bibinfo{year}{1985}), \bibinfo{pages}{1315}.
\newblock
Issue 6.
\showISSN{00129682}
\urldef\tempurl%
\url{https://doi.org/10.2307/1913210}
\showDOI{\tempurl}


\bibitem[Lamperti et~al\mbox{.}(2018)]%
        {lamperti2018agent}
\bibfield{author}{\bibinfo{person}{Francesco Lamperti}, \bibinfo{person}{Andrea Roventini}, {and} \bibinfo{person}{Amir Sani}.} \bibinfo{year}{2018}\natexlab{}.
\newblock \showarticletitle{Agent-based model calibration using machine learning surrogates}.
\newblock \bibinfo{journal}{\emph{Journal of Economic Dynamics and Control}}  \bibinfo{volume}{90} (\bibinfo{year}{2018}), \bibinfo{pages}{366--389}.
\newblock


\bibitem[Madhavan(2000)]%
        {madhavan2000market}
\bibfield{author}{\bibinfo{person}{Ananth Madhavan}.} \bibinfo{year}{2000}\natexlab{}.
\newblock \showarticletitle{Market microstructure: A survey}.
\newblock \bibinfo{journal}{\emph{Journal of financial markets}} \bibinfo{volume}{3}, \bibinfo{number}{3} (\bibinfo{year}{2000}), \bibinfo{pages}{205--258}.
\newblock


\bibitem[Majewski et~al\mbox{.}(2020)]%
        {majewski2020co}
\bibfield{author}{\bibinfo{person}{Adam~A Majewski}, \bibinfo{person}{Stefano Ciliberti}, {and} \bibinfo{person}{Jean-Philippe Bouchaud}.} \bibinfo{year}{2020}\natexlab{}.
\newblock \showarticletitle{Co-existence of trend and value in financial markets: Estimating an extended Chiarella model}.
\newblock \bibinfo{journal}{\emph{Journal of Economic Dynamics and Control}}  \bibinfo{volume}{112} (\bibinfo{year}{2020}), \bibinfo{pages}{103791}.
\newblock


\bibitem[McGroarty et~al\mbox{.}(2019a)]%
        {mcgroarty}
\bibfield{author}{\bibinfo{person}{Frank McGroarty}, \bibinfo{person}{Ash Booth}, \bibinfo{person}{Enrico Gerding}, {and} \bibinfo{person}{V.~L.~Raju Chinthalapati}.} \bibinfo{year}{2019}\natexlab{a}.
\newblock \showarticletitle{High frequency trading strategies, market fragility and price spikes: an agent based model perspective}.
\newblock \bibinfo{journal}{\emph{Annals of Operations Research}} \bibinfo{volume}{282}, \bibinfo{number}{1} (\bibinfo{year}{2019}), \bibinfo{pages}{217--244}.
\newblock
\showISBNx{1572-9338}


\bibitem[McGroarty et~al\mbox{.}(2019b)]%
        {mcgroarty2019high}
\bibfield{author}{\bibinfo{person}{Frank McGroarty}, \bibinfo{person}{Ash Booth}, \bibinfo{person}{Enrico Gerding}, {and} \bibinfo{person}{VL~Raju Chinthalapati}.} \bibinfo{year}{2019}\natexlab{b}.
\newblock \showarticletitle{High frequency trading strategies, market fragility and price spikes: an agent based model perspective}.
\newblock \bibinfo{journal}{\emph{Annals of Operations Research}} \bibinfo{volume}{282}, \bibinfo{number}{1-2} (\bibinfo{year}{2019}), \bibinfo{pages}{217--244}.
\newblock


\bibitem[Potters and Bouchaud(2003)]%
        {PottersBouchaud_statprop}
\bibfield{author}{\bibinfo{person}{Marc Potters} {and} \bibinfo{person}{Jean-Philippe Bouchaud}.} \bibinfo{year}{2003}\natexlab{}.
\newblock \showarticletitle{More statistical properties of order books and price impact}.
\newblock \bibinfo{journal}{\emph{Physica A: Statistical Mechanics and its Applications}} \bibinfo{volume}{324}, \bibinfo{number}{1} (\bibinfo{year}{2003}), \bibinfo{pages}{133--140}.
\newblock


\bibitem[Preis et~al\mbox{.}(2006)]%
        {preis2006multi}
\bibfield{author}{\bibinfo{person}{Tobias Preis}, \bibinfo{person}{Sebastian Golke}, \bibinfo{person}{Wolfgang Paul}, {and} \bibinfo{person}{Johannes~J Schneider}.} \bibinfo{year}{2006}\natexlab{}.
\newblock \showarticletitle{Multi-agent-based order book model of financial markets}.
\newblock \bibinfo{journal}{\emph{Europhysics Letters}} \bibinfo{volume}{75}, \bibinfo{number}{3} (\bibinfo{year}{2006}), \bibinfo{pages}{510}.
\newblock


\bibitem[Velu(2020)]%
        {velu2020algorithmic}
\bibfield{author}{\bibinfo{person}{Raja Velu}.} \bibinfo{year}{2020}\natexlab{}.
\newblock \bibinfo{booktitle}{\emph{Algorithmic trading and quantitative strategies}}.
\newblock \bibinfo{publisher}{CRC Press}.
\newblock


\bibitem[Vyetrenko et~al\mbox{.}(2020)]%
        {vyetrenko2020get}
\bibfield{author}{\bibinfo{person}{Svitlana Vyetrenko}, \bibinfo{person}{David Byrd}, \bibinfo{person}{Nick Petosa}, \bibinfo{person}{Mahmoud Mahfouz}, \bibinfo{person}{Danial Dervovic}, \bibinfo{person}{Manuela Veloso}, {and} \bibinfo{person}{Tucker Balch}.} \bibinfo{year}{2020}\natexlab{}.
\newblock \showarticletitle{Get real: Realism metrics for robust limit order book market simulations}. In \bibinfo{booktitle}{\emph{Proceedings of the First ACM International Conference on AI in Finance}}. \bibinfo{pages}{1--8}.
\newblock


\bibitem[Vytelingum et~al\mbox{.}(2008)]%
        {soton263234}
\bibfield{author}{\bibinfo{person}{Perukrishnen Vytelingum}, \bibinfo{person}{Dave Cliff}, {and} \bibinfo{person}{Nicholas~R Jennings}.} \bibinfo{year}{2008}\natexlab{}.
\newblock \showarticletitle{Strategic bidding in continuous double auctions}.
\newblock \bibinfo{journal}{\emph{Artificial Intelligence}} \bibinfo{volume}{172}, \bibinfo{number}{14} (\bibinfo{year}{2008}), \bibinfo{pages}{1700--1729}.
\newblock


\bibitem[Wang and Wellman(2017)]%
        {wang2017spoofing}
\bibfield{author}{\bibinfo{person}{Xintong Wang} {and} \bibinfo{person}{Michael~Paul Wellman}.} \bibinfo{year}{2017}\natexlab{}.
\newblock \showarticletitle{Spoofing the limit order book: An agent-based model}. In \bibinfo{booktitle}{\emph{Workshops at the Thirty-First AAAI Conference on Artificial Intelligence}}.
\newblock


\bibitem[Zhu et~al\mbox{.}(2023)]%
        {hkex}
\bibfield{author}{\bibinfo{person}{Dingqiu Zhu}, \bibinfo{person}{Richard Wise}, {and} \bibinfo{person}{Tao Chen}.} \bibinfo{year}{2023}\natexlab{}.
\newblock \showarticletitle{An innovative approach for optimizing CCP default management through Agent Based Modelling}.
\newblock \bibinfo{journal}{\emph{Journal of Risk Management}} (\bibinfo{date}{Oct} \bibinfo{year}{2023}).
\newblock


\end{thebibliography}



\end{document}